# Efficient High-Order Participation Factor Computation via Batch-Structured Tensor Contraction

Mahsa Sajjadi, *Student Member, IEEE,* Kaiyang Huang, *Student Member*, Kai Sun*, Fellow, IEEE*

*Abstract--* Participation factors (PFs) quantify the interaction between system modes and state variables, and they play a crucial role in various applications such as modal analysis, model reduction, and control design. With increasing system complexity, especially due to power electronic devices and renewable integration, the need for scalable and high-order nonlinear PF (NPF) computation has become more critical. This paper presents an efficient tensor-based method for calculating NPFs up to an arbitrary order. Traditional computation of PFs directly from normal form theory is computationally expensive—even for second-order PFs—and becomes infeasible for higher orders due to memory constraints. To address this, a tensor contraction–based approach is introduced that enables the calculation of high-order PFs using a batching strategy. The batch sizes are dynamically determined based on the available computational resources, allowing scalable and memory-efficient computation.

## I. Introduction

MODERN power systems are becoming increasingly nonlinear and stressed due to the high penetration of inverter-based resources (IBRs) and operation closer to stability limits, leading to growing challenges in stability and dynamic performance—particularly with sustained oscillations triggered by large disturbances. To effectively mitigate poorly damped—or even growing—oscillations, it is essential to identify the system components, such as generators or IBRs, that significantly participate in the oscillatory modes.

Participation Factors (PFs) quantify the relationship between system modes and state variables and are widely used in applications such as oscillatory stability, model reduction, and controller placement [1]-[4]. However, traditional linear PFs, derived from small-signal models, are often insufficient in stressed, nonlinear conditions near the onset of the unstable behavior —especially in systems with high IBR penetration. To address these limitations, nonlinear participation factors (NPFs) provide deep insight into nonlinear interactions, help identify highly participating areas and controller effects, and complement transient simulations to offer a more comprehensive understanding of system dynamics. By incorporating higher-order terms from normal form theory, NPFs can quantify the contributions of system states to modal behavior under nonlinear conditions. This is achieved through a nonlinear coordinate transformation and Taylor series expansion around the equilibrium, allowing modal contributions to be analyzed across different nonlinear orders. This provides deeper insight into phenomena such as modal coupling and resonance that linear methods may entirely miss [5]-[9]. NPFs have broad applicability, for example, in model reduction, higher-order NPFs help identify dynamically significant components more reliably than linear PFs, enabling reduced-order models that preserve accuracy even under nonlinear or near-resonance conditions. Similarly, in the context of controller placement, higher-order NPFs provide a means to evaluate how control actions influence nonlinear modal behavior. This allows for more informed decisions regarding the location and design of supplemental controllers, particularly in systems where nonlinear interactions are critical to performance [9].

Despite their theoretical strength, computing higher-order NPFs is challenging due to the exponential growth in terms with system size and expansion order, leading to significant memory and computational demands. Consequently, prior studies are limited to small systems and low-order cases. Moreover, existing normal form–based methods rely on deeply nested loops, resulting in high computation times and poor scalability for large-scale or high-order analysis. Ref. [10] introduced tensor contraction to accelerate NPF computation, but its reliance on nested loops and high memory demands limits its scalability, restricting it to small systems. In contrast, this paper proposes a fully vectorized tensor contraction approach combined with an adaptive batching strategy that adjusts batch sizes based on available computational resources. This enables scalable, memory-efficient and significantly faster computation, making high-order NPF analysis practical for large-scale systems.

## II. Tensor Contraction Approach for NPF

This section introduces a tensor contraction-based formulation of NPFs, the batching approach for high-order tensor computation, and the proposed vectorized NPF calculation. Note that the formulation of NPFs in this work assumes that the state matrix is diagonalizable. In cases involving non-diagonalizable repeated eigenvalues, this assumption no longer holds, and the formulation must be appropriately updated to account for the resulting Jordan block structure.

### A. NPFs by Tensor

The system's dynamics are governed by differential equations, expressed as a Taylor series expansion about the equilibrium point [9]:

This work was supported in part by NSF grant ECCS-2329924.
M. Sajjadi, K. Huang and K. Sun are with the Department of EECS, University of Tennessee, Knoxville, TN (Emails: msajjad1@vols.utk.edu, khuang12@vols.utk.edu, kaisun@utk.edu).

$$\frac{d}{dt}x^k = A1_r^k x^r + A2_{st}^k x^s x^t + \ldots + AN_{ab\ldots c}^k x^a x^b \ldots x^c + \ldots \quad (1)$$

where each *AN*-coefficient corresponds to a tensor of type (1, *N*): $A1_r^k$ is the $k^{th}$ row and $r^{th}$ column of the Jacobian matrix, i.e. the matrix *A* of the linearized model, $A2_{st}^k$ is the $s^{th}$ row and $t^{th}$ column in $k^{th}$ Hessian matrix and $AN_{ab\ldots c}^k$ is a $N^{th}$ order derivative tensor [10]. In the nonlinear normal form expansion, each state variable $x_k$ is expressed as a series of terms involving products of modal coordinates about $z_i = z_{i0} e^{\lambda_i t}$:

$$x_k = \sum_{i=1}^{n} \phi_{ki} z_i + \sum_{i=1}^{n} \phi_{ki} \sum_{p=1}^{n} \sum_{q=1}^{n} h_{pq}^i z_p z_q + \ldots + \sum_{i=1}^{n} \phi_{ki} \sum_{r=1}^{n} \ldots \sum_{v=1}^{n} h_{rs\ldots v}^i z_r z_s \ldots z_v + \ldots \quad (2)$$

where $\phi$ and $\psi$ are right and left eigenvectors, and the *h*-coefficients for *K*-th order terms correspond to (1, *K*) tensors in the form of $hN_{\alpha\beta\ldots\gamma}^r$ to eliminate the nonlinearity of the *K*-th order in the transformed *z*-coordinates after plugging (2) into (1). The expression of (2) in the tensor form becomes:

$$x^k = \phi_r^k z^r + \phi_r^k h2_{st}^r z^s z^t + \ldots + \phi_r^k hN_{\alpha\beta\ldots\gamma}^r z^\alpha z^\beta \ldots z^\gamma + \ldots \quad (3)$$

These transformation tensors can be known by solving a set of linear equations. If the nonlinearity of a specific order *K* is significantly more dominant than any lower-order nonlinearities, the corresponding (1, *K*) tensors can be approximated by:

$$h_{rs\ldots v}^i = \frac{\sum_{j=1}^{n}\sum_{\alpha=1}^{n}\ldots\sum_{\gamma=1}^{n} \psi_{ij}\, \alpha_{j,\alpha\beta\ldots\gamma}\, \phi_{\alpha r}\phi_{\beta s}\ldots\phi_{\gamma v}}{\lambda_r + \lambda_s + \ldots + \lambda_v - \lambda_i} \quad (4)$$

or equivalently,

$$hN_{\alpha\beta\ldots\gamma}^r = (DN)^{-1} AN_{ab\ldots c}^l \psi_l^r \phi_\beta^b \ldots \phi_\gamma^c \quad (5)$$

Since the computation of *h*-coefficients is required in all approaches for calculating NPFs, we use (5) uniformly to compare the three approaches in the rest of the paper to demonstrate the merits of the proposed approach. Based on the definition of NPFs, consider the initial excitation where only the $k^{th}$ state is perturbed, $x_k(0)=\delta^{k0}$, which equals 1 for the $k^{th}$ state and 0 otherwise. Substituting into (3) yields:

$$\begin{aligned} x^k(0) &= \phi_r^k z^r(0) + \phi_r^k h2_{st}^r z^s(0) z^t(0) + \ldots + \phi_r^k hN_{\alpha\beta\ldots\gamma}^r z^\alpha(0) z^\beta(0) \ldots z^\gamma(0) \\ &= pN1^k + pN2^k + \ldots + pNM^k + \ldots + pNN^k \end{aligned} \quad (6)$$

Equation (6) expresses the initial condition as a sum of nonlinear participation factors of different orders, capturing contribution of the single mode to multi-mode nonlinear interactions. $z(0)$ is the nonlinear initial condition in modal coordinates. For the single mode (linear mode), the NPF simplifies to:

$$pN1^k = \phi_i^k z^i(0) = \phi_i^k (\zeta^i - h2_{st}^i \zeta^s \zeta^t - \ldots - hN_{\alpha\beta\ldots\gamma}^i \zeta^\alpha \zeta^\beta \ldots \zeta^\gamma) \quad (7)$$

where $\zeta^i = \psi_{k0}^i \delta^{k0}$. If excitation magnitude, $\delta^{k0}$, is small enough, $z^i(0) \approx \bar{\zeta}^i$ and (7) reduces to linear PFs. For higher-order mode interactions, the NPF contribution for a mode can be expressed compactly as:

$$pNM^k = \phi_i^k \circ hN_{\alpha\beta\ldots\chi}^i \circ (\zeta^{*\alpha} \circ \zeta^{*\beta} \circ \ldots \circ \zeta^{*\chi}) \quad (8)$$

where

$$\zeta^{*\alpha} = \zeta^\alpha - h2_{ab}^\alpha \zeta^a \zeta^b - hN_{cd\ldots e}^\alpha \zeta^c \zeta^d \ldots \zeta^e \quad (9)$$

The $N^{th}$ order NPF is obtained by contacting the right eigenvector with the nonlinear transformation tensor and the Hadamard product of *M* modal interaction terms, each involving modal coordinate corrections up to order *N*. Note, NPFs vary with initial excitation, requiring recomputation for different magnitudes; however, the proposed tensor contraction and batching method enables efficient and scalable computation, unlike the traditional methods.

### B. Batching Approach for High-Order Tensor Computation

High-order tensor computations for NPFs can quickly exceed available RAM as system size or order increases. To address this, this work presents batching strategy that splits the computation into smaller, memory-efficient segments. The full tensor size scales as $n^N$, where *n* is the number of states and *N* is the tensor order. Multiplying by the data type size (e.g., 8 bytes for double) gives the total memory required; if it exceeds a predefined threshold, the computation is split into batches. The concept of batching used in this work follows the same general principle found in the literature to process large datasets but is customized for enabling efficient computation of NPF under memory constraints. While batching can also be applied to traditional NPF methods, they remain significantly slower due to their non-vectorized structure. The algorithm evaluates the full tensor and checks its memory requirements. If it exceeds the limit, outermost dimensions are reduced incrementally until the remaining dimensions fit in memory. It then determines batch size and count, computes each batch's tensor subset via numerical Taylor expansion, and stores the results. After batching, the vectorized tensor contraction (Section C) is applied within each batch to compute NPFs. The inputs to algorithm 1 include the order of the derivative tensor ($n_{dim}$), the number of system states ($n_s$), a memory usage limit per batch in GB ($M_{limit}$), the floating-point precision format (*DataType*), and the equilibrium point ($x^*$). The algorithm outputs a cell array of batched derivative tensors *AN* and a vector *BC* indicating the number of batches per dimension. This approach is memory-efficient, scalable to high-order tensors and large systems, and parallelizable.

---

**Algorithm 1: Batched $N^{th}$ order derivative tensor**

**Require:** $n_{dim}$, $n_s$, $M_{limit}$ [GB], *DataType*, $x^*$
**Ensure:** *AN*, *BC* (batch count per dimension)
**1. Initialization:**
2.   ◇ $dims = n_s \times Ones(1, n_{dim})$
3.   ◇ $BC = Ones(1, n_{dim})$, $TC = prod(dims)$, $B = Byte(DataType)$
4.   ◇ $M = TC \times B \times 1024^{-3}$   [GB]
**5. Compute Batch Size and Batch Count**
**6. while** $M > M_{limit}$
7.    ◇ $n_{dim} \leftarrow n_{dim} - 1$, $new\_dims \leftarrow n_s \times Ones(1, n_{dim})$
8.    ◇ $n_f \leftarrow prod(new\_dims)$, $M_f \leftarrow n_f \times B \times 1024^{-3}$
9.    **if** $M_f > M_{limit}$
10.      $BC(n_{dim}+1) \leftarrow n_s$
11.   **else**
12.      $E_{max} \leftarrow floor(\frac{M_{limit}}{M_f})$, $BC(n_{dim}+1) \leftarrow ceil(\frac{n_s}{E_{max}})$
13.   **end**
14.   ◇ $M \leftarrow M_f$
**15. End while**
**16. Generate and save batched high order derivative**
17.   ◇ $nb \leftarrow prod(BC)$
18.   **For** b=1: nb
19.     $S_z \leftarrow$ index range for each batch
20.     $AN\{b\} \leftarrow numericalTaylor(x^*, S_z)$
21.   **end**
**22. Return *AN*, *BC***

## C. Proposed Vectorized NPF Calculation

Traditional NPF methods based on normal form theory rely on explicit matrix operations and deeply nested summations, resulting in high computational overhead and poor scalability for large systems. The proposed Algorithm 2 overcomes these limitations with a fully vectorized, tensor-based framework for high-order computation. The inputs to the algorithm are $AN$, $\Phi$, $\Psi$, $\Lambda$, $n$, $N$, $\epsilon$, which represent the batched $N^{th}$ order derivative tensor, right and left eigenvectors, eigenvalues, the number of states, the desired tensor order, and a threshold parameter used to filter out weakly excited modes. Steps 1–4 of algorithm perform successive contractions of the $N^{th}$ order derivative tensor with right eigenvectors to obtain $CN$ as part of forming (5). Steps 5–9 compute the summation of eigenvalues to form resonance terms. Step 11 divides $CN$ to $DN$ to yield $hN$ in (5). Steps 12–15 construct the $\Psi p$ factors from the left eigenvectors corresponding to the initial modal condition factors $\zeta$ in (7)-(9). Steps 16-17 permute and reshape $hN$ to align them with $\zeta$. Steps 18-19 apply the Hadamard product between the contracted $\Psi p$ tensors corresponding to $\zeta^*$ and $hN$ in (8). Step 20 sums the contributions from all nonlinear order combinations to obtain the NPF. This algorithm significantly improves computational speed, enabling practical computation of high-order NPFs in large-scale systems.

| Algorithm 2: Tensor-Based NPF Computation |
|---|
| **Require:** $AN$, $\Phi$, $\Psi$, $\Lambda$, $n$, $N$, $\epsilon$ |
| **Ensure:** $NPF_N$ |
| 1. **Compute $CN$ tensor contraction** |
| 2.  ◦ $T_1 \leftarrow TensorCont(\Phi^T, AN)$ |
| 3.  ◦ $T_i \leftarrow TensorCont(T_{i-1}, \Phi)$, $i=2,...,N$ |
| 4.  ◦ $CN \leftarrow TensorCont(T_N, \Psi)/N!$ |
| 5. **Compute $DN$ tensor contraction** |
| 6.  ◦ $\lambda \leftarrow diag(\Lambda)$, $OnesTensor \leftarrow ones([n, ..., n]_{1 \times (N+1)})$ |
| 7.  ◦ $\gamma_i \leftarrow reshape\left(\lambda, [1,...,\overset{\text{Element ith}}{\widehat{n}},...,1]_{1 \times (N+1)}\right)$ |
| 8.  ◦ $D_i \leftarrow dot(OnesTensor, \gamma_i)$, $i=1,...,(N+1)$ |
| 9.  ◦ $DN \leftarrow -D1 + \sum_{i=2}^{N+1} D_i$, $DN(|DN|<\epsilon) \leftarrow \epsilon$ |
| 10. **Compute $hN$ tensor contraction** |
| 11.  ◦ $hN \leftarrow CN./DN$ |
| 12. **Compute $\Psi N$ tensor contraction** |
| 13.  ◦ $p_1, p_2, ..., p_N \leftarrow$ Mesh grid over $[1\ n]$ |
| 14.  ◦ Flatten $p_1, p_2, ..., p_N$ |
| 15.  ◦ $\Psi_{p_i} \leftarrow \Psi(p_i, :)$, $i = 1, 2, ... N$ |
| 16.  ◦ $\bar{h}N \leftarrow permute(hN, [1, N+1, N, ..., 2])$ |
| 17.  ◦ $\bar{h}N \leftarrow reshape(\bar{h}N, (n, n^N))$ |
| 18.  ◦ $\Psi_{prod} \leftarrow \Psi_{p_1} \circ \Psi_{p_2} \circ ... \circ \Psi_{p_N}$ |
| 19.  ◦ $\Psi N \leftarrow -TensorCont(\bar{h}N, \Psi_{prod})$ |
| 20.    $NPF_N \leftarrow \Phi \cdot (\Psi + \sum_{N=2} \Psi N)$ |
| 21. **Return** $NPF_N$ |

## III. NUMERICAL EXPERIMENT

This section presents numerical results using three methods: traditional NPF calculation based on normal form theory (T), the tensor contraction method from [10] (TC), and the proposed vectorized batching tensor contraction method (VBT). Case studies include a three-machine system (6 states), a 39-bus system (130 states), a partitioned Northeast Power Coordinating Council (NPCC) system (351 states) and large-scale Polish 2383 (4251 states). Fig. 1 presents second-, third-, and fourth-order NPF results for the three-machine system using the three methods under highly nonlinear conditions in Mode 1. The results show that NPF values vary significantly across different orders for some states, highlighting the importance of computing higher-order terms to accurately capture nonlinear modal behavior. Also, all three methods produce consistent results, with total RMSE values between VBT and T and TC on the order of $10^{-7}$, indicating negligible numerical differences and confirming the accuracy of the proposed approach. Table I shows the time cost comparison for the three methods in four case studies. As high-order derivative tensors grow exponentially in size (e.g., Hessian: $n^3$, third-order derivative: $n^4$, fourth-order: $n^5$), direct computation for larges systems becomes impractical without batching. Table I's empty entries mark computations that failed due to memory limits, while highlighted times show per-batch runtimes using the proposed batching strategy. The traditional method (T) is significantly slow, with runtimes exceeding a day even for small to moderate systems and becomes infeasible for large scale systems as Table I shows. In contrast, the proposed VBT method achieves superior speed and scalability, enabling high-order NPF computation for large-scale systems.

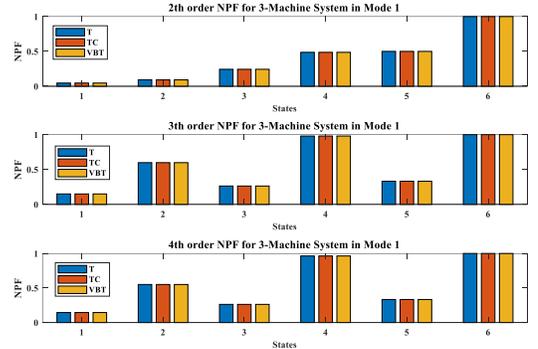

Fig. 1. $2^{th}$, $3^{th}$ and $4^{th}$ NPF in three methods for three-machine system

TABLE I
COMPARISON OF TIME COST FOR NPFs USING THREE APPROACHES

| Systems | Order | T | TC | VBT |
|---|---|---|---|---|
| Three-machine | $2^{th}$ | 0.0062 | 0.00139 | 0.000264 |
| | $3^{th}$ | 0.228 | 0.0103 | 0.000902 |
| | $4^{th}$ | 9.36 | 0.0142 | 0.00258 |
| IEEE 39-bus | $2^{th}$ | > 10 h | 1.27 | 0.108 s |
| | $3^{th}$ | > 24 h | 238 | 15.7 s |
| | $4^{th}$ | - | - | 254.59 s |
| NPCC | $2^{th}$ | > 24 h | 57.3 | 2.03 s |
| | $3^{th}$ | - | - | 332.83 s |
| | $4^{th}$ | - | - | 387.38 s |
| Polish 2383 | $2^{th}$ | - | - | 6.37 min |
| | $3^{th}$ | - | - | 918 min |
| | $4^{th}$ | - | - | 1040 min |

For the proposed approach, a memory limit of 8 GB per batch is imposed on a system with 15.73 GB of physical RAM in this study. This conservative limit accounts for the use of complex-valued data types, which require twice the memory of real-valued types, ensuring that batch computations remain feasible without exceeding available memory. Tables II and III show batching configurations and computational requirements for computing NPFs using the proposed algorithm. As shown, the total memory footprint increases rapidly with both system size and NPF order—reaching nearly 40 TB for the NPCC system at fourth order and up to 10.3 million TB for the Polish system at the same order. Without batching, such computations would be clearly infeasible on any current hardware platform due to memory constraints. However, by applying the proposed batching strategy, the memory usage per batch remains under 8 GB, making the computation feasible on a system with limited physical RAM. Note, the number of required batches for very large-scale systems becomes extremely large, highlighting the need for parallelization to manage computational efficiency. Also, focusing only on a subset of critical eigenvalues and their corresponding eigenvectors, which are most relevant for stability and control analysis, can further reduce the computational burden.

TABLE II
BATCHING CONFIGURATIONS AND COMPUTATIONAL REQUIREMENTS

| Systems | 39-bus (4$^{th}$ order) | NPCC (3$^{rd}$ order) | NPCC (4$^{th}$ order) |
|---|---|---|---|
| Total Size | 276.63 GB | 113.09 GB | 39,694 GB |
| Max Batch | [130, 130, 130, 130, 3] | [351, 351, 351, 24] | [351, 351, 351, 24, 1] |
| Batch Count | [1, 1, 1, 1, 44] | [1, 1, 1, 15] | [1, 1, 1, 15, 351] |
| # Batches | 44 | 15 | 5,265 |
| Per Batch | 6.38 GB | 7.73 GB | 7.73 GB |

TABLE III
BATCHING CONFIGURATIONS FOR LARGE-SCALE POLISH SYSTEM

| Systems | Polish 2383 (4251) | | |
|---|---|---|---|
| order | 2$^{th}$ order | 3$^{th}$ order | 4$^{th}$ order |
| Total size | 572.35 GB | 2.4e6 GB | 1.03e10 GB |
| Max Batch | [4251, 4251, 59] | [4251, 4251, 59, 1] | [4251, 4251, 59, 1, 1] |
| Batch Count | [1, 1, 73] | [1, 1, 73, 4251] | [1, 1, 73, 4251, 4251] |
| # Batches | 73 | 310323 | 1.32e9 |
| Per Batch | 7.94 GB | 7.94 GB | 7.94 GB |

## IV. CONCLUSION

This paper proposes a vectorized tensor contraction–based framework for computing high-order NPFs through an efficient batching strategy. The approach significantly reduces computational time and enables scalable computation for large-scale systems with limited hardware resources. By vectorizing the tensor operations and batching the data, the proposed method overcomes the memory bottlenecks inherent in traditional normal form–based and previous contraction-based NPF methods. Numerical results demonstrate the ability of the framework to compute NPFs up to arbitrary order for systems with thousands of state variables, where existing methods fail.